
\rightline{\bf ILG-TMP-93-03}
\smallskip
\rightline{\bf May, 1993}
\smallskip
\rightline{solv-int/9305004}
\bigskip\bigskip
\centerline{\bf Alexander A.Belov\footnote{$^{\#}$}
{Int. Inst. for Math. Geophys., Warshavskoe sh. 79, k.2 113556 Russia, e-mail:
mitpan\% adonis@sovam.com [Belov]}, Karen D.Chaltikian\footnote{$^{\$}$}
{Landau Inst. for Theor. Phys., Vorobyevskoe sh.2, Moscow 117334 Russia,
e-mail: chalt@cpd.landau.free.msk.su}}
\bigskip
\medskip
\centerline{\bf LATTICE VIRASORO FROM LATTICE KAC-MOODY}
\bigskip

{\sl  We propose a new version of quantum Miura transformation on the lattice
based on the lattice Kac-Moody algebra. In particular, we built Faddeev-
Takhtadjan-Volkov lattice Virasoro algebra from lattice current algebra
and discuss the possibility of existence of the lattice analogue of the
Sugawara construction.}
\vfill
\eject

\def\H#1{{\cal H}^{(#1)}}
\centerline{\bf 0. Introduction }
\bigskip

Lattice versions of the well-known continuous symmetries became of great
interest last time because of their possible relation to some certain models
of conformal matter coupled to 2D quantum gravity. The activity in this
direction began with the work of Faddeev and Takhtadjan [1]
on the clasical lattice
analogue of Liouville model, where for the first time classical lattice
version of the Virasoro algebra was obtained.
 It has been developed further
in the context of studying the integrable hierarchies related to matrix
models , which seem to be very relevant for the description
of the minimal matter coupled to gravity [2]. At the same time, the zoo
of the lattice symmetries is being expanded very fast last years, providing
a considerable amount of relevant material  both for physicists and
mathematicians
[3-5].

The aim of the present paper is to extend further the analogy between
the lattice world and the
continuous one, discussing  a possible lattice analogue of the well-known
Sugawara construction. In particular, starting from the lattice Kac-Moody
(LKM) algebra [6-8], we obtain the lattice Virasoro-
Faddev-Takhtadjan-Volkov (FTV)
algebra [1] in both quantum and classical cases.
\bigskip
\bigskip
\bigskip
\centerline{\bf 1. Lattice Kac-Moody algebra}
\bigskip

In this section we give the basic definitions and notations concerning the
LKM, following the work [8]. Let $R^{\pm}$ be the two solutions of the
Yang-Baxter equation
$$R_{12}^{\pm}R_{13}^{\pm}R_{23}^{\pm}=R_{23}^{\pm}R_{13}^{\pm}R_{12}^{\pm}
\eqno(1)$$
associated with the standard representation of $U_q(sl(N))$:
$$R^{+}=q^{1/2}\pmatrix{q^{-1}&0&0&0\cr 0&1&q^{-1}-q&0\cr 0&0&1&0\cr
0&0&0&q^{-1}\cr},\qquad \qquad
R^{-}=q^{-1/2}\pmatrix{q&0&0&0\cr 0&1&0&0\cr 0&q-q^{-1}&1&0\cr
0&0&0&q\cr}\eqno(2)$$

LKM-algebra ${\cal K}_N$ on the lattice with $N$ sites  is defined as a
free algebra of matrix elements of the matrix
$$J(n)\equiv \pmatrix{J(n)_{11}&J(n)_{12}\cr J(n)_{21}&J(n)_{22}\cr}
\quad \in \quad {\cal K}_N \otimes End({\bf C}^2)\eqno(3)$$
such that
$$J(n)_{11}J(n)_{22}-q^{-1}J(n)_{21}J(n)_{12}=q^{1/2}\eqno(4)$$
with the relations
$$\eqalign{J(n)_1J(n)_2&=R^{+}J(n)_2J(n)_1R^{-},\cr
J(n)_1J(n+1)_2&=J(n+1)_2{(R^{+})}^{-1}J(n)_1,\cr
J(n)_1J(m)_2&=J(m)_2J(n)_1 \qquad\hbox{for}\;\;m\neq n-1,\; n,\;n+1\cr}
\eqno(5)$$
Lattice analogues of quantum vertex operators are defined via the relations
[8]:
$$h(n+1)=J(n)h(n)\eqno(6)$$
and form the representation of LKM as
$$\eqalign{&J(n)_1R^{-}h(n)_2=h(n)_2J(n)_1\cr &
J(n)_1h(n+1)_2=R^{+}h(n+1)_2J(n)_1\cr}\eqno(7)$$
Further on we consider an infinite lattice in order to avoid possible
problems with boundary conditions.

In the classical limit $q\to 1$ one obtains the following Poisson brackets
for the currents
$$\eqalign{&\{J(n)_1,J(n)_2\}=r^{+}J(n)_1J(n)_2+J(n)_1J(n)_2r^{-},\cr
&\{J(n)_1,J(n+1)_2\}=-J(n+1)_2r^{+}J(n)_1,\cr
&\{J(n)_1,J(n-1)_2\}=-J(n)_1r^{-}J(n-1)_2,\cr}$$
where classical $r^{\pm}$-matrices defined from the
expansion of $R^{\pm}$ at $k\to \infty$ as $R^{\pm}=1+{1\over i}{\pi \over 2k}
r^{\pm}$ have the form:
$$r^{+}=\pmatrix{1&0&0&0\cr 0&-1&4&0\cr 0&0&-1&0\cr 0&0&0&1\cr}
\qquad\qquad  r^{-}=\pmatrix{-1&0&0&0\cr 0&1&0&0\cr 0&-4&1&0\cr 0&0&0&-1\cr}
$$
Classical continuous limit is defined as follows $(x\equiv n\Delta )$
$$\eqalign{&J_n^{11}\to 1-\Delta j_{11}(x)+\ldots ,\qquad\qquad
J_n^{12}\to -\Delta j_{12}(x)+\ldots,\cr
&J_n^{21}\to \Delta j_{21}(x)+\ldots\qquad\qquad
J_n^{22}\to 1-\Delta j_{22}(x)+\ldots\cr}\eqno(8)$$
with the condition (4) transformed to
$$j_{11}(x)+j_{22}(x)=0$$
The fields $j$ form classical $sl(2)$-KM algebra
$$\eqalign{&\{j_{11}(x),j_{21}(y)\}=2j_{21}(x)\delta (x-y)\cr
&\{j_{11}(x),j_{12}(y)\}=-2j_{12}(x)\delta (x-y)\cr
&\{j_{11}(x),j_{21}(y)\}=2\delta '(x-y)+4j_{11}(x)\delta (x-y)\cr
&\{j_{11}(x),j_{11}(y)\}=-\delta '(x-y)\cr}\eqno(9)$$
\bigskip
\bigskip
\bigskip
\centerline{\bf 2. Lattice Virasoro from Lattice Kac-Moody.}
\bigskip

Recall first the standard $Lu(1)$ realization of the FTV algebra [1]. If one
has $Lu(1)$-current with the exchange relations
$$U_nU_{n+1}=q^2U_{n+1}U_n$$
then the quantum Miura transformation gives us the FTV-generators in the
following form [3,4]
$$\sigma_n=1+U_n^{-1}+U_{n+1}+q^{-1}U_n^{-1}U_{n+1}\eqno(10)$$
So defined lattice fields $\sigma_n$ form quantum FTV algebra [3,4],
which will be written below.
Classical version of this construction starts from the Poisson brackets
$$\{u_n,u_m\}=u_nu_m(\delta_{n,m+1}-\delta_{m,n+1})$$
The Miura transformation
$$S_n={u_{n+1}\over (1+u_n)(1+u_{n+1})}\eqno(11)$$
leads to the FTV algebra
$$\eqalign{&\{S_n,S_{n+1}\}=S_nS_{n+1}(1-S_n-S_{n+1})\cr &
\{S_n,S_{n+2}\}=-S_nS_{n+1}S_{n+2}\cr}\eqno(12)$$
\bigskip
\bigskip
\bigskip
\leftline{\qquad \qquad \it 2.a Quantum case.}
\medskip

In this section we give explicit formulas, expressing the generators of the
quantum FTV algebra  in terms of the LKM generators. Namely, define
the operators
$$F_n=J_n^{12}J_n^{21},\qquad\qquad M_n={J_n^{12}J_{n+1}^{21}\over
 J_n^{22}J_{n+1}^{22}}\eqno(13)$$
Direct calculation provides us with the following quantum commutators
$$\eqalign{&[F_n,F_{n+1}]=(1-q^{-2})F_{n+1}M_n^{-1}F_n\cr &
[M_n,M_{n+1}]=q^{3/2}(q^2-1)M_nF_{n+1}^{-1}M_{n+1}\cr}\qquad \qquad
\eqalign{&[F_n,M_n]_{q^{-1}}=(q-q^{-1})(F_n+q^{3/2}M_n)\cr &
[F_{n+1},M_n]_q=-(q-q^{-1})(F_{n+1}+q^{3/2}M_n)\cr}\eqno(14)$$
in which the reader acquainted with the abovecited literature on the
subject readily recognizes quantum FTV algebra after the identification
$$\sigma_{2n}\equiv\;F_n\qquad \qquad \sigma_{2n+1}=M_n\eqno(15)$$
Thus, some new type of quantum Miura transformation is defined, dealing not
with lattice $u(1)$ but $sl(2)$ currents.
Some final remarks about the representation of the algebra above
should be made.
Direct calculation show, that although the Jacobi identity is satisfied in the
sector ($Vir$, $Vir$, $h$), where $h$ is quantum vertex defined in (6)-(7)
no good relation like (6) exist for the fields $F$ and $M$. Nonetheless,
one can easily check that $LKM$ currents (3) form the quantum
representation of the FTV algebra in $F-M$-realization
$$\eqalign{&F_nJ_{n-1}^{11}=J_{n-1}^{11}F_n-(1-q^{-2})F_nM_{n-1}^{-1}F_{n-1}
{(F_{n-1}+q)}^{-1} J_{n-1}^{11}\cr &
F_nJ_{n-1}^{12}=J_{n-1}^{12}F_n-(1-q^{-2})F_nM_{n-1}^{-1} J_{n-1}^{12}\cr &
F_nJ_{n}^{12}=q^2J_{n}^{12}F_n+q^{3/2}(q^2-1)J_n^{12}\cr &
F_nJ_{n}^{21}=q^{-2}J_{n}^{21}F_n+q^{3/2}(q^{-2}-1)J_n^{21}\cr &
F_nJ_{n+1}^{11}=J_{n+1}^{11}F_n+(1-q^{-2})J_{n+1}^{11}{(F_{n+1}+q)}^{-1}
F_{n+1}M_n^{-1}F_n\cr &
F_nJ_{n+1}^{21}=J_{n+1}^{21}F_n+(1-q^{-2})J_{n+1}^{21}{F_{n+1}}^{-1}M_n\cr
&M_nJ_n^{11}=q^{-2}J_n^{11}M_n-(1-q^{-2})J_n^{11}F_n{(F_n+q)}^{-1}
\cr &M_nJ_n^{12}=q^{-2}J_n^{12}M_n-(1-q^{-2})J_n^{12}\cr &
M_nJ_n^{21}=J_n^{21}M_n-q^{3/2}(q^2-1)J_n^{21}F_n^{-1}M_n\cr &
M_nJ_{n+1}^{11}=q^2J_{n+1}^{11}M_n+(q^2-1)J_{n+1}^{11}F_{n+1}{(F_{n+1}+q)}^{-1}
\cr & M_nJ_{n+1}^{12}=J_{n+1}^{12}M_n+q^{3/2}(q^2-1)M_nF_n^{-1}J_{n+1}^{12}\cr
&M_nJ_{n+1}^{21}=q^2J_{n+1}^{21}M_n+(q^2-1)J_{n+1}^{21}\cr}\eqno(16)$$

\bigskip
\bigskip
\bigskip
\leftline{\qquad \qquad \it 2.b Classical limit.}
\medskip

In this paragraph we study the classical limit of the construction above.
First of all, we give the Poisson brackets version of (14).
$$\eqalign{&\{F_n,F_{n+1}\}=F_nF_{n+1}{1\over M_n}\cr
&\{M_n,M_{n+1}\}=M_nM_{n+1}{1\over F_{n+1}}\cr
&\{F_n,M_n\}=F_nM_n(1+{1\over F_n}+{1\over M_n})\cr
&\{F_{n+1},M_n\}=-F_{n+1}M_n(1+{1\over F_{n+1}}+{1\over M_n})\cr}\eqno(17)$$
what means that FTV generators themselves are defined as
$$A_{2n}\equiv {1\over F_n},\qquad \qquad A_{2n+1}\equiv {1\over M_n}$$
and form the original FTV-algebra (12).

Using (8) we find in continuous classical limit
$$\eqalign{&F(x)=-T^{\hbox{root}}(x)=-j_{12}(x)j_{21}(x)\cr
&A_n\to -{1\over T^{\hbox{root}}(x)}+\ldots\cr}\eqno(18)$$
This allows us to interpret the abovementioned FTV algebra
in $F-M$ realization  in terms of certain integrable model. Namely,
 consider the following pair of Poisson brackets:
$${\{m_n,f_n\}}_1=m_nf_n,\qquad\qquad {\{m_n,f_{n+1}\}}_1=-m_nf_{n+1}
\eqno(19)$$
$$\eqalign{&{\{f_n,f_{n+1}\}}_2=-m_nf_nf_{n+1},\qquad \qquad
{\{m_n,f_{n+1}\}}_2=
-m_nf_{n+1}(m_n+f_{n+1})\cr
&{\{m_n,m_{n+1}\}}_2=-m_nm_{n+1}f_{n+1},\qquad{\{f_n,m_n\}}_2=-
f_nm_n(f_n+m_n)\cr}\eqno(20)$$

It should be mentioned that the first bracket may be interpreted
as induced one {\it via} the following construction
$$m_n=e^{\phi_n-\phi_{n+1}},\qquad f_n=e^{p_n}$$
where $\phi_n$ and $p_n$ have canonical bracket
$$\{\phi_n,p_n\}=1$$
The variables $F_n, M_n$ can be thought about as those inverse to
$f_n, m_n$ respectively. Then the algebra FTV (12) is obtained as the
sum of the first and the second brackets introduced above. This pair of
brackets generates by means of bihamiltonian procedure
an infinite series of integrals in involution
$$\eqalign{&\H{1}=\sum f_n+m_n\cr &
\H{2}=\sum {f_n^2\over 2}+f_nm_n+f_{n+1}m_n+{m_n^2\over 2}\cr &
\ldots \qquad \ldots \qquad \ldots\cr}\eqno(21)$$
 for the Hamiltonians
$${\H0}_m=\sum_n\ln m_n \qquad \hbox{or} \qquad {\H0}_f=
\sum_n\ln f_n$$
Note that after substituting the explicit form of $f_n$ and $m_n$ in (21)
and taking continuous limit via the rule (8) one obtains expressions
for the hamiltonians, containing the inverse powers of the $sl(2)$-currents.
\bigskip
\bigskip
\bigskip
\centerline{\bf 3. A hint for Sugawara construction: {\it pro} and
{\it contra}.}
\bigskip

In this section we discuss various aspects of the intriguing problem of
building
a lattice analogue of the Sugawara construction. First of all, the very concept
of the
{\it lattice analogue} seems to be not very well defined. In the absence of
any more
clear definition\footnote{$^{\#}$}{Say geometrical one, like in
continuous WZW model} ,
 one could use the continuous limit of the FTV generator as a starting point.
However, even in the most trivial case of $Lu(1)$ Miura transformation (11) the
continuous limit of the FTV generator (12) is not the Sugawara energy-momentum
tensor $T_{u(1)-\hbox{Sug}}=u^2(x)$ but some twisted one
$T(x)=u^2(x)+\alpha u'(x)$ where $\alpha \neq 0$.

To understand this specifically lattice difficulties one should recall that
{\it latticization} is a kind of $q$-deformation [9]. General philosophy of
quantization is in that generically {\it splits the degeneracy}. In a wide
sense it implies
that some objects coinciding  in continuous limit may become different
on the lattice.
 For example, the well-known isomorphism between bosons and fermions in
two dimensions dissapears after the "latticization".
As a more advanced example
one can consider the direct correspondence between $N=2$ superconformal
theories and topological ones. In continuous case this correspondence bases on
existence of the {\it twist } operation. However, no satisfactory
analogue of such
an operation exists to connect these two classes of the models on the
lattice [10].
To our conjecture, the continuous Sugawara construction is an example of
certain degeneracy. It is well-known, that any Sugawara energy-momentum
tensor can be
represented as a sum of mutually commuting parts, corresponding to
the appropriate
subalgebras and coset-spases. This, in particular, takes place in
the framework of
the bosonization procedure or BRST quantization. Such a decomposition in the
continuous limit is a relevant technical tool necessary for better
understanding
the structure of the theory.
The situation is drastically different on the lattice.
Given a number of mutually commuting FTV algebras there is no any local
mapping of the type
$${\cal T}:\qquad {\hbox{FTV}}_1\otimes {\hbox{FTV}}_2\otimes\ldots
{\hbox{FTV}}_N \to \hbox{FTV}$$
or in other words there is no  any function of the form
$$\eqalign{ S_n\; &=\cr &=\;
{\cal T}(S_{n+a_1}^{(1)},\ldots ,S_{n+a_{k_1}}^{(1)}\; \vert
S_{n+b_1}^{(2)},\ldots ,S_{n+b_{k_2}}^{(2)}\; \vert \ldots
\vert S_{n+c_1}^{(N)},\ldots ,
S_{n+c_{k_N}}^{(N)})\cr }$$
where all the sets $\{ a_i\}$, $\{ b_i\}$,$\ldots$, $\{ c_i\}$ {\bf finite}.
Thus the conformal properties of different subsystems cannot be unified.
In the case of  $Lsl(2)$ current algebra one can built three
mutually commuting FTV algebras but due to the argument above
there is no any FTV algebra for the whole system of the lattice currents.
These FTV generators and their continuous limits on the classical level
have the form
$$\eqalign{& S_n^{(1)}={u_{n+1}\over p_np_{n+1}}\quad \to\quad
I^2(x)+I'(x)\cr &
 S_n^{(2)}={u_{n+1}\over (1+u_n)(1+u_{n+1})}\quad \to\quad
(j_{12}j_{21})(x)-I^2(x)-I'(x)\cr &
S_n^{(3)}={h_{n+1}\over (1+h_n)(1+h_{n+1})}\quad \to\quad
{j_{22}(x)}^2+{j_{22}}^{\prime}(x)\cr}$$
where
$$\eqalign{& u_n={F_{n-1}\over M_{n-1}}\cr &
p_n=1-F_n+{F_n\over M_n}\cr &
h_n=J_n^{22}\cr &
I(x)=(\log j_{21}(x))'(x)\cr}$$

More generally it can be shown that for each LKM algebra there
is a set of $N$ mutually commuting FTV algebras where $N$ is  dimension
of the LKM. In continuous limit the sum of all these FTV-fields turns
out to be the (twisted) Sugawara energy-momentum
tensor. Thus, the specific property of lattice conformal models is that the
number of lattice fields necessary to describe their conformal
properties is always equal to the number of the fields parametrizing
the phase space of the model. Lattice WZW theory is a particular example
of this situation. Another example advocating for the validity of this
conjecture is from lattice $W$-algebras, recently obtained by the present
authors [5]. Namely, for any pair of integer numbers  $N_1<N_2$ one cannot
extract any
$LW_{N_1}$-subalgebra from the $LW_{N_2}$-algebra.

 Of course, there is an alternative approach to the lattice
conformal invariance based on the notion of $L$-operator, probably more
fruitfull in the case of lattice WZW-model than the direct one, based
on the analysis of FTV-algebras for constituent parts of the model. This
approach is another part of the program in progress.

\vfill
\eject
\centerline{\bf R E F E R E N C E S}
\bigskip
\bigskip
{\obeylines\smallskip
1. L. Faddeev, L. Takhtadjan, Lect. Notes in Phys. {\bf 246} (1986), 166
\smallskip
2. S. Kharchev et al., Nucl. Phys. {\bf B366} (1991), 569
\smallskip
\qquad L. Bonora, C. Xiong, {\it 'Multimatrix models without continuous
limit"},
\qquad preprint {\bf SISSA-ISAS 211/92/EP} (1992)
\smallskip
3. O. Babelon, Phys. Lett. {\bf B238} (1990), 234
\smallskip
\qquad O. Babelon, L. Bonora, Phys. Lett. {\bf B253} (1991), 365
\smallskip
\qquad L. Bonora, V. Bonservizi, Nucl. Phys. {\bf B390} (1993), 205
\smallskip
4. A. Volkov, Zap.Nauch.Sem.LOMI {\bf 150} (1986), 17; ibid. {\bf 151} (1987)
24;
\qquad Theor. Math. Phys. {\bf 74} (1988) 135; Phys. Lett. {\bf A167} (1992),
345
\smallskip
5. A. Belov, K. Chaltikian, {\it "Lattice analogues of W-algebras and Classical
Integrable Equations"},
\qquad preprint {\bf ILG-TMP-93-01} (1993), hep-th/9303166, to appear in
Phys.Lett.{\bf B}
\smallskip
\qquad A. Belov, K. Chaltikian, {\it "Lattice analogue of W-infinity and
Discrete KP-hierarchy"},
\qquad preprint {\bf ILG-TMP-93-02} (1993)
6. N. Reshetikhin, M. Semenov-Tian-Shansky, Lett. Math. Phys. {\bf 19} (1990),
133
\smallskip
7. A. Alekseev, L. Faddeev, M. Semenov-Tian-Shansky, Lect. Notes in Math. {\bf
1510} (1992), 148
\smallskip
8. F. Falceto, K. Gawedzki, {\it "Lattice Wess-Zumino-Witten models and Quantum
Groups"},
\qquad preprint {\bf IHES/P/92/73} (1992)
\smallskip
9. A. Belov, K. Chaltikian, {\it "Q-deformations of Virasoro algebra and
lattice Conformal Theories"},
\qquad preprint {\bf ILG-TMP-92-04} (1992), to appear in Mod.Phys.Lett.{\bf A}
\smallskip
10. A. Belov, K. Chaltikian, {\it "Lattice Analogues of N=1,2 Superconformal
Theories}, in preparation
\smallskip}
\end